\documentstyle[prl,aps]{revtex}
\begin{document}

Pis'ma v ZhETF {\bf 66}, 627 (1997)

\vskip 4mm

\centerline{\large \bf Combined effect of nonmagnetic and magnetic
scatterers}
\centerline{\large \bf on critical temperatures of superconductors}
\centerline{\large \bf with different gap anisotropy}

\vskip 2mm

\centerline{Leonid A. Openov}

\vskip 2mm

\centerline{\it Moscow State Engineering Physics Institute
(Technical University)}
\centerline{\it 115409 Moscow, Russia}
\centerline{E-mail: opn@supercon.mephi.ru}

\vskip 4mm

\begin{quotation}

The combined effect of nonmagnetic and magnetic defects and impurities on
critical temperatures of superconductors with different gap anisotropy is
studied theoretically within the weak coupling limit of the BCS model. An
expression is derived which relates the critical temperature to relaxation
rates of charge carriers by nonmagnetic and magnetic scatterers, as well as
to the coefficient of anisotropy of the superconducting order parameter on
the Fermi surface. Particular cases of $d$-wave, $(s+d)$-wave, and
anisotropic $s$-wave superconductors are briefly discussed.

\end{quotation}

\vskip 4mm

This paper is motivated by conflicting experimental results concerning the
symmetry of the superconducting order parameter $\Delta(\bf p)$ in
high-temperature superconductors (HTSCs) and the suppression of the critical
temperature $T_c$ of HTSCs by defects and impurities. Indeed, while the
majority (though not all) of experiments support the $d$-wave
superconductivity in HTSCs \cite{d-wave}, the observed degradation of $T_c$
by impurities or radiation-induced defects \cite{experiment} is more gradual
than predicted theoretically for $d$-wave superconductors \cite{theory}.

To resolve this contradiction, a number of suggestions have been made,
including anisotropic $s$-wave symmetry of $\Delta(\bf p)$ \cite{Openov},
momentum dependence of impurity scattering \cite{Haran}, strong coupling
effects resulting in crossover from Cooper pairs to local bosons
\cite {Sadovskii}, {\it etc}. Note, however, that theoretical analysis of
$T_c$ degradation by defects and impurities is usually restricted to the
specific case of spin-independent scattering potential, i.e., to the case of
nonmagnetic scatterers only. Meanwhile a lot of experiments give evidence for
the presence of magnetic scatterers (along with nonmagnetic ones) in
non-stoichiometric HTSCs, e.g., in oxygen-deficient, doped or irradiated
samples \cite{magnetic}.

The goal of this paper is to work out a theoretical framework for a
description of {\it combined} effect of nonmagnetic and magnetic scatterers
on $T_c$ of a superconductor with anisotropic $\Delta(\bf p)$ (in what
concerns an isotropic $s$-wave superconductor, its $T_c$ is insensitive to
nonmagnetic defects \cite{Anderson}, while the $T_c$ suppression by magnetic
defects is given by a well-known Abrikosov-Gor'kov theory \cite{Abrikosov}).
We use the weak coupling limit of the BCS model for superconducting pairing
and the Born approximation for impurity scattering. In what follows, we do
not specify the microscopic mechanism of superconductivity. We set
$\hbar =k_B=1$ throughout the paper.

The Hamiltonian of a superconductor containing both nonmagnetic and magnetic
scatterers reads
\begin{equation}
\hat{H}=\sum_{{\bf p},\sigma}\xi ({\bf p})\hat{a}^{+}_{{\bf p}\sigma}%
\hat{a}^{}_{{\bf p}\sigma}~+\sum_{{\bf p},{\bf p}^\prime,\sigma,\sigma^%
\prime}U({\bf p},\sigma;{\bf p}^\prime,\sigma^\prime)\hat{a}^{+}_{{\bf p}%
\sigma}\hat{a}^{}_{{\bf p}^\prime\sigma^\prime}~+\sum_{{\bf p},{\bf p}^%
\prime}V({\bf p},{\bf p}^\prime)\hat{a}^{+}_{{\bf p}\uparrow}\hat{a}^{+}_%
{-{\bf p}\downarrow}\hat{a}^{}_{-{\bf p}^\prime\downarrow}\hat{a}^{}_%
{{\bf p}^\prime\uparrow},
\label{Hamiltonian}
\end{equation}
where $\xi({\bf p})=\epsilon({\bf p})-\mu$ is the quasiparticle energy
measured from the chemical potential,
$U({\bf p},\sigma; {\bf p}^\prime,\sigma^\prime)$ is the matrix element for
electron scattering by randomly distributed impurities (defects) from the
state $({\bf p}^\prime,\sigma^\prime)$ to the state $({\bf p},\sigma)$, and
$V({\bf p},{\bf p}^\prime)$ is the BCS pair potential.

We assume for simplicity that electron scattering is isotropic in the
momentum space, the amplitude of the scattering by an isolated nonmagnetic
(magnetic) scatterer being $u_n$ ($u_m$). Then the relaxation times $\tau_n$
and $\tau_m$ are given by the standard "golden rule" formulas
\begin{equation}
\frac{1}{\tau_n}=2\pi c_n|u_n|^2N(0)~,~~\frac{1}{\tau_m}=2\pi c_m|u_m|^2N(0),
\label{tau_nm}
\end{equation}
where $c_n$ and $c_m$ are the concentrations of scatterers, $N(0)$ is the
density of electron states at the Fermi level. Note that the commonly
accepted expression for $|u_m|^2$ is $J^2S(S+1)$/4, where $J$ is the energy
of electron-impurity exchange interaction, $S$ is the impurity spin.

In order to account for anisotropy of the superconducting state, we assume a
factorizable pairing interaction of the form \cite{Abrikosov2}
\begin{equation}
V({\bf p},{\bf p}^\prime)=-V_0\phi({\bf n})\phi({\bf n}^\prime),
\label{V(p,p')}
\end{equation}
where ${\bf n}={{\bf p}}/p$ is a unit vector along the momentum. Then the
order parameter $\Delta ({\bf p})$ is \cite{Abrikosov2}
\begin{equation}
\Delta ({\bf p})=\Delta_0\phi({\bf n}),
\label{Delta}
\end{equation}
where $\Delta_0$ depends on the temperature. Thus the function
$\phi({\bf n})$ specifies the anisotropy of $\Delta ({\bf p})$ in the
momentum space ($\phi({\bf n})\equiv 1$ for isotropic pairing). The
self-consistent equation for $\Delta ({\bf p})$ can be derived by means of
Green's functions technique (see, e.g., \cite {Abrikosov}). It is as follows:
\begin{equation}
\Delta({\bf p})=-\sum_{{\bf p}^\prime}V({\bf p},{\bf p}^\prime)\langle%
\hat{a}^{}_{-{\bf p}^\prime\downarrow}\hat{a}^{}_{{\bf p}^\prime\uparrow}%
\rangle=-T\sum_{\omega}\sum_{{\bf p}^\prime}V({\bf p},{\bf p}^\prime)%
\frac{\Delta_\omega({\bf p}^\prime)}{\omega^{\prime 2}+\xi^2({\bf p}^\prime)%
+|\Delta_\omega({\bf p}^\prime)|^2},
\label{Eq_Delta}
\end{equation}
where $\omega =\pi T(2n+1)$ are Matsubara frequencies, and the equations for
$\Delta_\omega({\bf p})$ and $\omega^\prime$ are
\begin{equation}
\Delta_\omega({\bf p})=\Delta({\bf p})+(c_n|u_n|^2-c_m|u_m|^2)\sum_{{\bf p}^%
\prime}\frac{\Delta_\omega({\bf p}^\prime)}{\omega^{\prime 2}+\xi^2({\bf p}^%
\prime)+|\Delta_\omega({\bf p}^\prime)|^2},
\label{Eq_Delta_omega}
\end{equation}
\begin{equation}
\omega^\prime=\omega-i(c_n|u_n|^2+c_m|u_m|^2)\sum_{{\bf p}^\prime}%
\frac{i\omega^\prime+\xi({\bf p}^\prime)}{\omega^{\prime 2}+\xi^2({\bf p}%
^\prime)+|\Delta_\omega({\bf p}^\prime)|^2}.
\label{Eq_omega'}
\end{equation}

Since $\Delta({\bf p})=0$ at $T=T_c$, in the case $T\rightarrow T_c$ we have
from (\ref{Eq_Delta_omega}), (\ref{Eq_omega'}), taking (\ref{tau_nm}) into
account:
\begin{equation}
\Delta_\omega({\bf p})=\Delta({\bf p})+\frac{1}{2|\omega^\prime|}\left(%
1/\tau_n-1/\tau_m\right)\langle\Delta_\omega({\bf p})\rangle,
\label{Eq_Delta_omega2}
\end{equation}
\begin{equation}
\omega^\prime=\omega+\frac{1}{2}\left(1/\tau_n+1/\tau_m%
\right)sign(\omega),
\label{Eq_omega'2}
\end{equation}
where angular brackets $\langle ... \rangle$ stand for the average over the
Fermi surface (FS):
\begin{equation}
\langle ... \rangle=\int_{FS}(...)\frac{d\Omega_{\bf p}}{|\partial%
\xi({\bf p})/\partial {\bf p}|}\Biggl/\int_{FS}\frac{d\Omega_%
{\bf p}}{|\partial\xi({\bf p})/\partial {\bf p}|}.
\label{average}
\end{equation}

Substituting (\ref{Eq_Delta_omega2}) and (\ref{Eq_omega'2}) in
(\ref{Eq_Delta}) and taking (\ref{V(p,p')}) into account, we have after
rather simple but time consuming algebraic transformations:
\begin{equation}
\ln\left(\frac{T_{c0}}{T_c}\right)=\pi T_c\sum_{\omega}\frac{1}{|\omega|+%
\frac{1}{2}\left(1/\tau_n+1/\tau_m\right)}\left[\frac{1}{2|\omega|}%
\left(1/\tau_n+1/\tau_m\right)-\frac{\langle\phi({\bf n})\rangle^2}%
{\langle\phi^2({\bf n})\rangle}\cdot\frac{1/\tau_n-1/\tau_m}{2\left%
(|\omega|+1/\tau_m\right)}\right].
\label{Tc}
\end{equation}
Here $T_{c0}$ is the critical temperature in the absence of impurities and
defects (at $1/\tau_n=1/\tau_m=0$). At this stage it is convenient to
introduce the coefficient $\chi$ of anisotropy of the order parameter on the
FS \cite{Abrikosov2}, \cite{Openov}
\begin{equation}
\chi=1-\frac{\langle\phi({\bf n})\rangle^2}{\langle\phi^2({\bf n})\rangle}=%
1-\frac{\langle\Delta({\bf p})\rangle^2}{\langle\Delta^2({\bf p})\rangle}.
\label{chi}
\end{equation}
For isotropic $s$-wave pairing we have $\Delta({\bf p})\equiv const$ on the
FS; therefore, $\langle\Delta({\bf p})\rangle^2=\langle\Delta^2({\bf p})%
\rangle$, and $\chi=0$. For a superconductor with $d$-wave pairing we have
$\chi=1$ since $\langle\Delta({\bf p})\rangle=0$. The range $0<\chi<1$
corresponds to anisotropic $s$-wave pairing or to mixed $(d+s)$-wave pairing.
The higher is the anisotropy of $\Delta({\bf p})$ (e.g., the greater is the
partial weight of a $d$-wave in the case of mixed pairing), the closer to
unity is the value of $\chi$.

Making use of the definition (\ref{chi}) and the formula
\cite{Prudnikov}
\begin{equation}
\sum_{k=0}^{\infty}\left(\frac{1}{k+x}-\frac{1}{k+y}\right)=\Psi(y)-\Psi(x),
\label{digamma}
\end{equation}
where $\Psi$ is the digamma function, we obtain from (\ref{Tc}):
\begin{equation}
\ln\left(\frac{T_{c0}}{T_c}\right)=(1-\chi)\left[\Psi\left(\frac{1}{2}+%
\frac{1}{2\pi T_c\tau_m}\right)-\Psi\left(\frac{1}{2}\right)\right]+%
\chi\left[\Psi\left(\frac{1}{2}+\frac{1}{4\pi T_c}\cdot\left(\frac{1}{\tau_n}%
+\frac{1}{\tau_m}\right)\right)-\Psi\left(\frac{1}{2}\right)\right].
\label{Tc2}
\end{equation}
In two particular cases of ($i$) both nonmagnetic and magnetic scattering in
an isotropic $s$-wave superconductor ($\chi =0$) and ($ii$) nonmagnetic
scattering only in a superconductor with arbitrary anisotropy of
$\Delta({\bf p})$ ($1/\tau_m=0$, $0\leq\chi\leq 1$), the Eq.(\ref{Tc2})
reduces to well-known expressions \cite{Abrikosov}, \cite{Abrikosov2}
\begin{equation}
\ln\left(\frac{T_{c0}}{T_c}\right)=\Psi\left(\frac{1}{2}+%
\frac{1}{2\pi T_c\tau_m}\right)-\Psi\left(\frac{1}{2}\right)
\label{Tc3}
\end{equation}
and
\begin{equation}
\ln\left(\frac{T_{c0}}{T_c}\right)=\chi\left[\Psi\left(\frac{1}{2}+%
\frac{1}{4\pi T_c\tau_n}\right)-\Psi\left(\frac{1}{2}\right)\right].
\label{Tc4}
\end{equation}
respectively.

The Eq. (\ref{Tc2}) is obviously more general than Eqs.(\ref{Tc3}) and
(\ref{Tc4}) which are commonly used for the analysis of experimental data on
$T_c$ suppression by defects and impurities in HTSCs \cite{experiment2}. In
fact, making use of Eq. (\ref{Tc3}) {\it or} Eq.(\ref{Tc4}) one {\it assumes
a priori} that either ($i$) the order parameter in HTSCs is isotropic in the
momentum space or ($ii$) the magnetic scatterers in HTSCs are completely
absent. The latter assumption is often supplemented with a speculation about
pure $d$-wave symmetry of $\Delta({\bf p})$ \cite {experiment3} (i.e., one
intentionally restricts himself to the case $\chi=1$ instead of attempts to
extract the value of $\chi$ from the experiment). In our opinion, the
experimental dependencies of $T_c$ versus impurity (defect) concentration or
radiation dose should be analyzed within the framework of the theory
presented above, see Eq.(\ref{Tc2}). One should not {\it guess} as to the
degree of anisotropy of $\Delta({\bf p})$ and the type of scatterers, but try
to {\it determine} the value of {$\chi$} and relative weights of magnetic and
nonmagnetic components in electron scattering through comparison of
theoretical predictions with available or specially performed experiments.

Now let us consider the limiting cases of weak and strong scattering
($T_{c0}-T_c<<T_{c0}$ and $T_c\rightarrow 0$ respectively). At
$1/4\pi T_{c0}\tau_n<<1$ and $1/4\pi T_{c0}\tau_m<<1$ (weak scattering) one
has from (\ref{Tc2}):
\begin{equation}
T_{c0}-T_c\approx\frac{\pi}{4}\left[\frac{\chi}{2\tau_n}+%
\frac{1-\chi/2}{\tau_m}\right].
\label{Tc5}
\end{equation}
In particular cases ($i$) and ($ii$) considered above, Eq.(\ref{Tc5})
reduces to well-known expressions \cite {experiment2}
\begin{equation}
T_{c0}-T_c\approx\frac{\pi}{4\tau_m}
\label{Tc6}
\end{equation}
and
\begin{equation}
T_{c0}-T_c\approx\frac{\pi\chi}{8\tau_n}
\label{Tc7}
\end{equation}
for initial $T_c$ reduction by magnetic (at $\chi=0$) or nonmagnetic (at
arbitrary value of $\chi$) scatterers respectively.

In what concerns the strong scattering limit, we recall that in the BCS
theory, nonmagnetic scattering alone is insufficient for the not-$d$-wave
superconductivity ($0\leq\chi<1$) to be destroyed completely
\cite {Abrikosov2}; at $1/\tau_m=0$, the value of $T_c$ asymptotically goes
to zero as $1/\tau_n$ increases (whereas $T_c$ of a $d$-wave superconductor
with $\chi=1$ vanishes at a critical value $1/\tau_n^c=\pi T_{c0}/\gamma%
\approx 1.764T_{c0}$, where $\gamma=e^C\approx 1.781$, $C$ is the Eiler
constant). On the other hand, magnetic scattering in the absence of
nonmagnetic one ($1/\tau_n=0$) is known to suppress the isotropic $s$-wave
superconductivity with $\chi=0$ at a critical value
$1/\tau_m^c=\pi T_{c0}/2\gamma\approx 0.882T_{c0}$ \cite {Abrikosov}.

Based on the Eq.(\ref{Tc2}), it is straightforward to derive the general
condition for impurity (defect) suppression of $T_c$ of a superconductor
having an arbitrary anisotropy coefficient $\chi$ and containing both
nonmagnetic and magnetic scatterers:
\begin{equation}
\frac{1}{\tau_{eff}^c}=\frac{\pi}{\gamma}2^{\chi-1}T_{c0},
\label{tau_eff_c}
\end{equation}
where $\tau_{eff}^c$ is the critical value of the effective relaxation time
$\tau_{eff}$ defined as
\begin{equation}
\frac{1}{\tau_{eff}}=\left(\frac{1}{\tau_m}\right)^{1-\chi}\cdot%
\left(\frac{1}{\tau_n}+\frac{1}{\tau_m}\right)^{\chi}.
\label{tau_eff}
\end{equation}

From Eqs. (\ref{tau_eff_c}) and (\ref{tau_eff}) one can see that
$1/\tau_{eff}^c$ increases monotonically with both $1/\tau_n$ and $1/\tau_m$
at any value of $\chi$, with the exception of the case $\chi=0$ when
$1/\tau_{eff}$ doesn't depend on $1/\tau_n$, see (\ref{tau_eff}). If $\chi$
is close to unity (strongly anisotropic $\Delta(\bf p)$), then
$1/\tau_{eff}\approx 1/\tau_n+1/\tau_m$, i.e., the contribution of
nonmagnetic and magnetic scattering to pair breaking is about the same. If
$\chi<<1$ (almost isotropic $\Delta(\bf p)$), then
$1/\tau_{eff}\approx 1/\tau_m$, i.e., $\tau_{eff}$ is determined primarily by
magnetic scattering. The higher is the anisotropy coefficient $\chi$, the
greater is the relative contribution of nonmagnetic scatterers to $T_c$
degradation with respect to magnetic ones. If nonmagnetic scattering is
absent ($1/\tau_n=0$), then $1/\tau_{eff}=1/\tau_m$ at any value of $\chi$.

We note however that while the concept of the effective relaxation time
$\tau_{eff}$ can be used for evaluation of the {\it critical} level of
nonmagnetic and magnetic disorder, it is not possible to express $T_c$ in
terms of $\tau_{eff}$ in the {\it whole range} $0\leq T_c\leq T_{c0}$, see
(\ref{Tc2}). In other words, the combined effect of nonmagnetic and magnetic
scattering on $T_c$ cannot be described by a single universal parameter
depending on the values of $\tau_n$, $\tau_m$, and $\chi$. For example,
$1/\tau_{eff}=0$ at $1/\tau_m=0$ and $0\leq\chi<1$ no matter what the value
of $1/\tau_n$ is. On the one hand, as follows from (\ref{tau_eff_c}), the
zero value of $1/\tau_{eff}^c$ in this case points to
the fact that in a BCS superconductor with not-$d$-wave symmetry of
$\Delta(\bf p)$ the critical level of disorder cannot be reached in the
absence of magnetic scattering, in accordance with \cite{Abrikosov2}. On the
other hand, the zero value of $1/\tau_{eff}$ obviously doesn't imply that
$T_c$ of a not-$d$-wave superconductor is completely insensitive to
nonmagnetic scatterers at $1/\tau_m=0$ and $0<\chi<1$, see (\ref{Tc2}).
Hence, while the quantity $1/\tau_{eff}^c$ characterizes the critical
strength of impurity (defect) scattering corresponding to $T_c=0$, the
quantity $1/\tau_{eff}$ (when it is less than $1/\tau_{eff}^c$) doesn't
determine the value of $T_c$ unequivocally.

Based on Eqs. (\ref{tau_eff_c}) and (\ref{tau_eff}), it is possible to derive
the following expression for the critical value of $1/\tau_n$ in the presence
of magnetic scattering:
\begin{equation}
\frac{1}{\tau_n^c}=\frac{1}{\tau_m}\left[2\left(\frac{\pi T_{c0} \tau_m}%
{2\gamma}\right)^{1/\chi}-1\right].
\label{tau_n_c}
\end{equation}
This expression is valid as long as $1/\tau_m<\pi 2^{\chi-1}T_{c0}/\gamma$
since otherwise the superconductivity is completely suppressed solely by
magnetic impurities. The value of $1/\tau_n^c$ decreases as $1/\tau_m$
increases at constant $\chi$ or as $\chi$ increases at constant $1/\tau_m$.
The finite value of $1/\tau_n^c$ in the presence of magnetic scatterers could
reconcile the experimentally observed disorder-induced suppression of
$T_c$ of HTSCs below 4.2K \cite{experiment} with theories of not purely
$d$-wave symmetry of $\Delta(\bf p)$ in HTSCs, e.g., anisotropic $s$-wave
symmetry or mixed $(d+s)$-wave symmetry.

In conclusion, the results obtained provide the basis for evaluation of
the degree of anisotropy of the superconducting order parameter (and hence
its possible symmetry) as well as the type of scatterers (magnetic or
nonmagnetic) in high-$T_c$ superconductors through careful comparison of
theoretical predictions with the experiments on impurity-induced and
radiation-induced reduction of the critical temperature. We hope that the
present paper will serve as a stimulus for experiments on combined effect of
nonmagnetic and magnetic scattering in the copper-oxide superconductors.

This work was supported by the Russian Foundation for Fundamental Research
under Grant No 97-02-16187. The author would like to thank V.F.Elesin and
A.V.Krasheninnikov for fruitful discussions at the early stage of this work
and V.A.Kashurnikov for useful comments.

\vskip 4mm

{\it Note added in proof}. After submission of this paper I became aware of a
similar study by A.A.Golubov and I.I.Mazin [Phys. Rev. B {\bf 55}, 15146
(1997)] which generalizes Abrikosov-Gor'kov solution to the case of a
multiband superconductor with interband order parameter anisotropy.


\vskip 4mm

\begin{references}

\bibitem{d-wave}J.Annett, N.Goldenfeld, and A.J.Leggett, in {\it Physical
Properties of High Temperature Superconductors}, edited by D.M.Ginsberg
(World Scientific, Singapore, 1996), Vol.5, and references therein.
\bibitem{experiment}A.G.Sun, L.M.Paulius, D.A.Gajewski {\it et al.}, Phys.
Rev. B {\bf 50}, 3266 (1994); J.Giapintzakis, D.M.Ginsberg, M.A.Kirk, and
S.Ockers, Phys. Rev. B {\bf 50}, 15967 (1994); F.Rullier-Albenque, A.Legris,
H.Berger, and L.Forro, Physica C {\bf 254}, 88 (1995); S.K.Tolpygo, J.-Y.Lin,
M.Gurvitch {\it et al.}, Phys. Rev. B {\bf 53}, 12454 (1996); V.F.Elesin,
K.E.Kon'kov, A.V.Krasheninnikov, and L.A.Openov, Zh. Eksp. Teor. Fiz.
{\bf 110}, 731 (1996).
\bibitem{theory}R.J.Radtke, K.Levin, H.-B.Sch\"uttler, and M.R.Norman, Phys.
Rev. B {\bf 48}, 653 (1993); K.Levin, Y.Zha, R.J.Radtke {\it et al.},
J. Supercond. {\bf 7}, 563 (1994).
\bibitem{Openov}L.A.Openov, V.F.Elesin, and A.V.Krasheninnikov, Physica C
{\bf 257}, 53 (1996).
\bibitem{Haran}G.Hara\'n and A.D.S.Nagi, Phys. Rev. B {\bf 54}, 15463 (1996).
\bibitem{Sadovskii}M.V.Sadovskii and A.I.Posazhennikova, Pis'ma v Zh. Eksp.
Teor. Fiz. {\bf 65}, 258 (1997).
\bibitem{magnetic}A.M.Finkel'stein, V.E.Kataev, E.F.Kukovitskii,
G.B.Teitel'baum, Physica C {\bf 168}, 370 (1990); G.Xiao, M.Z.Cieplak,
J.Q.Xiao, and C.L.Chien, Phys. Rev. B {\bf 42}, 8752 (1990); B. vom Hedt,
W.Lisseck, K.Westerholt, and H.Bach, Phys. Rev. B {\bf 49}, 9898 (1994);
J.Axn\"as, W.Holm, Yu.Eltsev, and \"O.Rapp, Phys. Rev. B {\bf 53}, 3003
(1996); V.P.S.Awana, D.A.Landinez, J.M.Ferreira {\it et al.}, Mod. Phys.
Lett. B {\bf 10}, 619 (1996); T.G.Togonidze, V.N.Kopylov, N.N.Kolesnikov,
and I.F.Schegolev, Czech. J. Phys. {\bf 46}, 1379 (1996).
\bibitem{Anderson}P.W.Anderson, J. Phys. Chem. Solids {\bf 11}, 26 (1959).
\bibitem{Abrikosov}A.A.Abrikosov and L.P.Gor'kov, Zh. Eksp. Teor. Fiz.
{\bf 39}, 1781 (1960).
\bibitem{Abrikosov2}A.A.Abrikosov, Physica C {\bf 214}, 107 (1993).
\bibitem{Prudnikov}A.P.Prudnikov, Yu.A.Brychkov, and O.I.Marichev,
Integraly i Ryady. - M.: Nauka, 1981, p.655.
\bibitem{experiment2}P.S.Prabhu, M.S.R.Rao, U.V.Varadaraju, and G.V.S.Rao,
Phys. Rev. B {\bf 50}, 6929 (1994); E.M.Jackson, B.D.Weaver, G.P.Summers
{\it et al.}, Phys. Rev. Lett. {\bf 74}, 3033 (1995); S.K.Tolpygo, Phys. Rev.
Lett. {\bf 75}, 3197 (1995); P.Monod, K.Maki, and F.Rullier-Albenque, Phys.
Rev. Lett. {\bf 75}, 3198 (1995); T.Kluge, Y.Koike, A.Fujiwara {\it et al.},
Phys. Rev. B {\bf 52}, 727 (1995); M.Brinkmann, H.Bach, and K.Westerholt,
Phys. Rev. B {\bf 54}, 6680 (1996); B.D.Weaver, G.P.Summers, R.L.Greene
{\it et al.}, Physica C {\bf 261}, 229 (1996).
\bibitem{experiment3}J.-Y.Lin, H.D.Yang, S.K.Tolpygo, and M.Gurvitch,
Czech. J. Phys. {\bf 46}, 1187 (1996); D.M.Ginsberg, J.Giapintzakis, and
M.A.Kirk, Czech. J. Phys. {\bf 46}, 1203 (1996).

\end{references}
\end{document}